\documentclass[aps,pra,groupedaddress,amsmath,amssymb]{revtex4-1}
\usepackage{mathtools}
\usepackage{graphicx,subcaption}  
\usepackage[font=small,  justification=justified, singlelinecheck=false,
   format=plain]{caption} 
\usepackage{dcolumn}   
\usepackage{bm}        
\usepackage{verbatim}   
\usepackage{wrapfig}
\usepackage{float, array}
\usepackage{amsmath}
 \usepackage{amssymb}
\usepackage{verbatim, url}
\usepackage{fullpage}
\usepackage{enumerate}
\usepackage{bm}
\usepackage{notes2bib}
\usepackage{lpic}
\usepackage{pgfplots}
\usepackage[T1]{fontenc}
\usepackage[latin1]{inputenc}
\usepackage[extra]{tipa}
\usepackage{accents}
\usepackage{epstopdf}
\usepackage{appendix}
 \usepackage{amsthm}

\newlength{\dhatheight}

\newcommand{\nn}{\nonumber}
\newcommand{\f}[2] {\frac{#1}{#2}}

\newcommand{\beq}{\begin{equation}}
\newcommand{\eeq}{\end{equation}}
\newcommand{\beqn}{\begin{eqnarray}}
\newcommand{\eeqn}{\end{eqnarray}}

\DeclarePairedDelimiter\abs{\lvert}{\rvert}%
\DeclarePairedDelimiter\norm{\lVert}{\rVert}%

\makeatletter
\let\oldabs\abs
\def\abs{\@ifstar{\oldabs}{\oldabs*}}
\let\oldnorm\norm
\def\norm{\@ifstar{\oldnorm}{\oldnorm*}}
\makeatother

\begin{document}

\title{Oscillator bath simulation of the spin bath}
\author{Seyyed M.H. Halataei}
\affiliation{
   Department of Physics, University of Illinois at Urbana-Champaign,\\
   1110 West Green Street, Urbana, Illinois 61801, USA
   }
   
\date{Dec. 14, 2017}
\begin{abstract}
Open quantum systems are subject to interaction with their surrounding environment. In many applications, at low temperatures, quantum environments fall into two universality classes of models: Caldeira-Leggett oscillator bath models and Prokof'ev-Stamp spin bath models. The two classes are commonly recognized to be distinct and to have strikingly different effects on principal systems, except at weak coupling limits. However, I show here, in contrast, that oscillator bath models can simulate the effect of spin bath models in strong coupling limit of the spin bath. I choose parameters of the oscillator bath models such that they produce incoherent relaxation, just as in the spin bath models, with relaxation rates for a two-state system (a qubit) equivalent to those of the spin bath models in the strong coupling limit.
\end{abstract}

\maketitle

\section{Introduction}
Quantum theory was originally developed in the context of \emph{isolated} microscopic systems whose interactions with their environments were negligible (e.g. the atoms in a beam). The theory was tested successfully in this domain in the early twentieth century and some of its founding fathers, such as Niels Bohr, believed that it would not be applicable in a larger domain where systems are strongly coupled to complex environments \cite{Leggett1989,BohrCollected}.

Advancement of experimental techniques and equipments in recent decades, however, showed that quantum theory does apply in a broader range. It can well describe behaviors of \emph{open} quantum systems, as large as a few microns, that are strongly coupled to their complex \emph{environments} (e.g. the phase of the Cooper pairs, in SQUIDs, that is coupled to phonons, radiation field, normal electrons, nuclear spins, etc.) \cite{Caldeira83, Leggett87, Friedman2000, vanderWal2000, Makhlin2001, Clarke2008, Oliver2011, Oliver2016}.        

In handling complex environments coupled to open quantum systems, the \emph{effects} that the environments exert on the principal systems are of the main interest, not the behaviors of the environments in their own right. As a result, theorists attempt to model complex environments by mapping them onto simpler ones that are better tractable and have the same \emph{effects} on the principal systems. Two of these simple models that are now well established in the literature are the oscillator bath and spin bath models \cite{Caldeira83, Leggett87, Leggett1989, Prokofev1998, Prokofev93, Tupitsyn97, Prokofev2000, Prokofev95, Prokofev95b, Schlosshauer2007, Weiss2008}. 

The oscillator bath model consists of a set of non-interacting simple harmonic oscillators that are individually coupled to the principal system. Caldeira and Leggett \cite[p. 439]{Caldeira83} showed that at absolute zero temperature, any arbitrary environment whose each degree of freedom is only weakly perturbed, by the principal system, can be mapped onto an oscillator bath. 

It is important to note that although each degree of freedom of the \emph{environment} is weakly perturbed in this model, the \emph{principal system} can be strongly perturbed by the overall effect of all the oscillators. We use the term {\it weak coupling limit} to refer to the case in which each degree of freedom of the environment is weakly perturbed, and the term {\it strong coupling limit} when they are strongly perturbed. However, in both cases the principal system can be weakly or strongly perturbed.  

The oscillator bath model has been extensively used in the literature to model phonons, electrons, magnons, spinons, holons, quasiparticles, etc. at low energies and temperatures  \cite{Leggett87, Nesi2007, Thorwart2004, Grifoni1999, Nesi2007b, Schlosshauer2007, Weiss2008,Eckern84}. 

The spin bath model, on the other hand, consists of microscopic spins that are independently coupled to the principal system. In real scenarios, these spins usually interact with one another weakly \cite{Prokofev2000, Gatteschi2006}. The spin bath model can be studied at both weak and strong coupling limits and there are practical cases associated to each of them \cite{Prokofev1998, Prokofev93, Tupitsyn97, Prokofev2000, Prokofev95, Prokofev95b, Gatteschi2006, Garg2013, Garg2015, Wernsdorfer1999, Takahashi2011, Wernsdorfer2002, Garg2009}. 

Caldeira, Neto and de Carvalho \cite{Caldeira93} demonstrated that the effect of a non-interacting spin bath, in the weak coupling limit, on a principal system can be simulated by an oscillator bath whose spectral density function is suitably chosen \cite{Caldeira93}. 

Weiss \cite{Weiss2008} obtained the same result by use of the fluctuation-dissipation theorem for each degree of freedom of the environment, which is permissible in the weak coupling limit (Sec. 3.5 and 6.1 of \cite{Weiss2008}). 

Despite this success in simulating noninteracting spin bath in the weak coupling limit by an oscillator bath, the scheme was not extended  to the case of self interacting spin bath, mainly due to the difficulty of calculating spin correlation function for each spin in a self interacting environment.

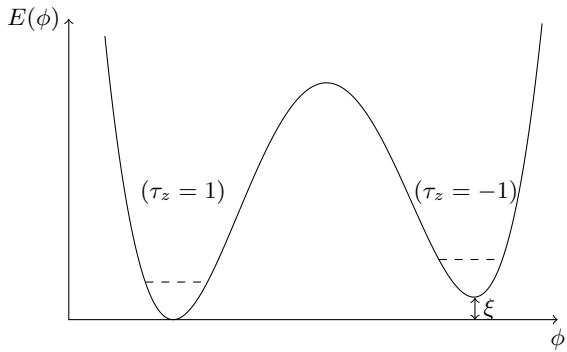
\begin{figure}[htbp]
\centering
\begin{tikzpicture}[xscale=2]

\draw [<->] (-2.5+.8,4) node[left] {$E(\phi)$} -- (-2.5+.8,0)  -- (2.5-.95,0) node[below] {$\phi$};
\draw [domain=-1.46:1.45, samples=300] plot (\x,{3* (\x*\x-1)*(\x*\x-1) + .3/2*(\x+1)});
\draw [dashed] (-1.19,1/2) -- (-.778,1/2) ;
\node at (-.94,1.7) {$(\tau_z = 1)$};
\node at (.94,1.7) {$(\tau_z = -1)$};

\draw [dashed] (0.7598,1/2+.3) -- (1.18193,1/2+.3);


\draw [<->] (1,0) -- (1,0.3/2-.012/2) node[right] {$\xi$} -- (1,0.3-.012);

\end{tikzpicture}
\centering
\captionsetup{singlelinecheck = false, justification=justified}
\caption{Double well potential or energy landscape $E(\phi)$ \cite{energycomment} of the high energy Hamiltonian of the principal system. $\xi$ is the bias energy difference between the two minima \cite{xicomment} and $\phi$ can be multidimensional. The dashed lines represents the energy of the localized states in {\it absence} of tunneling.}
\label{Fig1}
\end{figure}

A direct solution of the spin bath problem was reached by Prokof'ev and Stamp \cite{Prokofev1998, Prokofev93, Tupitsyn97, Prokofev2000, Prokofev95, Prokofev95b}. They found that in the strong coupling limit the effect of an interacting spin bath on an effectively two-state system is to relax the system incoherently. More precisely, they obtained that under most conditions the principal system undergoes an incoherent relaxation with relaxation rate $\Gamma(\xi)$ that decreases by the increase of the bias energy difference $\xi$ between the two minima of the double well energy landscape $E(\phi)$ of the principal system (Fig. \ref{Fig1}). 

In regard to the simulation of the spin bath by the oscillator bath, Prokof'ev and Stamp, however, believed that the effect of a spin bath, in the strong coupling limit, is in complete contrast to the effect of an oscillator bath: ``All of this is in complete contrast to how inelastic tunneling works in the presence of an oscillator bath; there the relaxation rate typically increases as one moves away from resonance'' (i.e. as the bias energy increases) \cite{Prokofev2000}. They, hence, concluded that the spin bath model in the strong coupling limit has completely different effects on principal systems and is not comparable with the oscillator bath model \cite{Prokofev1998, Prokofev93, Tupitsyn97, Prokofev2000, Prokofev95, Prokofev95b}.  

In this paper, we, however, show that a Caldeira-Leggett oscillator bath model can simulate the effect of a spin bath in the strong coupling limit of the spin bath. We demonstrate that an oscillator bath can cause an incoherent relaxation with a relaxation rate $\Gamma(\xi)$ that decreases as one moves away from resonance(i.e. as the bias energy $\xi$ increases). By choosing an appropriate spectral density function $J(\omega)$ for the oscillator bath we obtain a relaxation rate $\Gamma(\xi)$ that is quantitatively comparable with that of the spin bath model. 

The existence of such simulation implies that for most practical purposes and as long as the effect of the environment in terms of its relaxation or decoherence rate is concerned, which is the case in practical applications of the spin bath \cite{Prokofev1998}, there is far less distinction between the spin bath and the oscillator bath models than has been previously recognized. 

The organization of this paper is as follows: In Secs. \ref{SpinBath}-\ref{OscillatorBath} we briefly review the spin bath and oscillator bath models, respectively. In Sec. \ref{Sim} we present the simulation of the spin bath model, in the strong coupling limit, by the oscillator bath model. We conclude in Sec. \ref{con}. 

\section{Spin Bath Model} \label{SpinBath}
The spin bath model \cite{Prokofev1998, Prokofev93, Tupitsyn97, Prokofev2000, Prokofev95, Prokofev95b} concerns the effect of the interaction of an effectively two-state system with an environment composed of microscopic spins (called spin bath). The principal system has usually multiple (or infinite) energy levels, but in low temperatures it acts as a two-state system (or a qubit) since only the two lowest lying states are occupied. Thus, one may truncate the high energy Hamiltonian of the system to obtain the effective two-state Hamiltonian 
\beq \label{HS}
H_S =  - \f{\Delta}{2} \hat{\tau}_x - \f{\xi}{2} \hat{\tau}_z,
\eeq
for the principal system (we set $\hbar = 1$ in this paper). When the high energy Hamiltonian has a double well potential or energy landscape \cite{energycomment}, $\Delta$ is interpreted as the tunneling matrix element between the two localized states in either wells, and $\xi$ is the bias energy difference between the two minima\cite{xicomment} (Fig. \ref{Fig1}). In Eq. \eqref{HS}, $\hat{\tau}_x$ and $\hat{\tau}_z$ are Pauli matrices and we assumed that the localized states in either wells are eigenstates of $\hat{\tau}_z$. Thus, $\hat{\tau}_x$ takes the system from a localized state in one well to the other localized state in the other well.  

The energy levels of the lowest localized states in {\it absence} of tunneling is drawn by dashed lines in Fig. \ref{Fig1}. The system is called to be at {\it resonance} when these levels coincide. The examples are $\xi =0$ or when $\xi$ is such that the energy level of a higher state in one well lines up with that of the lowest state in the other well.  

\begin{figure}[htbp]
\centering
\begin{tikzpicture}[xscale=1.5]
\draw [<->] (-2.7+.7,4) node[left] {$E(\phi)$} -- (-2.7+.7,0) -- (2.7-.4,0) node[below] {$\phi$};
\draw [domain=-1.46:1.45, samples=300] plot (\x,{3* (\x*\x-1)*(\x*\x-1) + .3/2*(\x+1)});
\draw [very thin,black] (-2+.4,1/2+.3/2-.2) node [below right] {$E_{+}$} -- (1.9-.4,1/2+.3/2-.2) ;
\draw [very thin,black] (-2+.4,1/2+.3/2+.2)  node [above right] {$E_{-}$} -- (1.9-.4,1/2+.3/2+.2);
\draw [<->] (+1.8+.2-.4,1/2+.3/2+.22) -- (+1.8+.2-.4,1/2+.3/2) -- node [right] {$\Delta E$} (+1.8+.2-.4,1/2+.3/2-.22) ;
\draw [<->] (1,0) -- (1,0.3/2-.012/2) node[right] {$\xi$} -- (1,0.3-.012);

\end{tikzpicture}
\caption{Double well potential or energy landscape $E(\phi)$ of the principal system. The tunneling effect lifts the degeneracy in the energy levels. $E_{\pm}$ denote the ground state and first excited state energies of the system with splitting $\Delta E$. $\xi$ denotes  the bias energy difference between the minima. For zero bias, $\xi=0$, the energy splitting and the tunneling matrix element becomes equal, $\Delta E = \Delta \ne 0$. For $\xi \lesssim \Delta$ the eigenstates of the Hamiltonian are localized in both wells.}
\label{Fig2}
\end{figure}
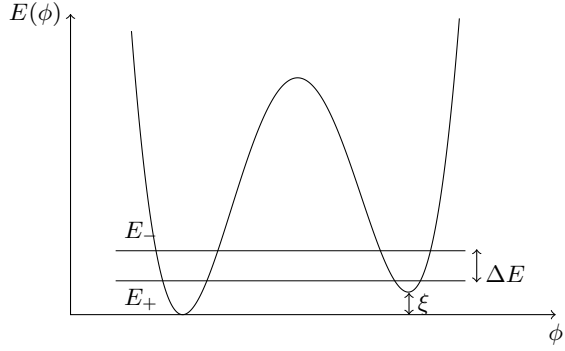

In reality, however, the tunneling effect lifts the degeneracy such that near or at resonance the energy eigenstates become localized in both wells and the difference between energy eigenvalues becomes finite (Fig. \ref{Fig2})
  
The surrounding microscopic spins in the environment usually have a nonzero self-Hamiltonian of the following generic form
\beq
H_{B,\text{spin}} = \sum_{k,k'=1}^N \sum_{\alpha, \beta=1}^3 V_{kk'}^{\alpha \beta} \hat{\sigma}_k^\alpha \hat{\sigma}_{k'}^\beta
\eeq
where $\hat{\vec{\sigma}}_k$ is the spin operator of the kth spin in the bath and $V_{kk'}^{\alpha \beta}$ are the interspin couplings. These spins can interact with the system if they are close enough to it. One of the effects of interaction is to cause the bias energy $\xi$ to fluctuate by an amount $\xi_0$. The latter plays an important role in the relaxation rate of the system caused by the spin bath as we later see (Eqs. (\ref{Gamma}-\ref{Gamma'})). The full dynamics of the system can be found by calculating the truncated Hamiltonian of the system plus bath (the Universe), as is done by Prokof'ev and Stamp, for a typical interaction Hamiltonian between the system and bath. They found that the universe truncated Hamiltonian is \cite{Prokofev2000},
\begin{align} 
\nn H_{U,\text{spin}} &=  - \f{\Delta}{2}\left\{\hat{\tau}_- \cos \left[\Phi - i \sum_{k=1}^N \vec{\alpha}_k \cdot \hat{\vec{\sigma}}_k \right] + \text{H.c.} \right\} \\ 
\nn &- \f{\xi}{2} \hat{\tau}_z + \hat{\tau}_z \sum_{k=1}^N \vec{\omega}_k^\parallel \cdot \hat{\vec{\sigma}}_k + \sum_{k=1}^N \vec{\omega}_k^\perp \cdot \hat{\vec{\sigma}}_k  \\ \label{Hps}
&+ \sum_{k,k'=1}^N \sum_{\alpha, \beta=1}^3 V_{kk'}^{\alpha \beta} \hat{\sigma}_k^\alpha \hat{\sigma}_{k'}^\beta,
\end{align}
where $\hat{\vec{\tau}}$ are Pauli operators of the principal system, as before, and $\Phi$, $\vec{\alpha}_k$, $\vec{\omega}_k^\parallel$ and $\vec{\omega}_k^\perp$ are the parameters of the model that depend on the high energy Hamiltonian of the principal system (See Ref. \cite{Prokofev2000}). The term in the curly brackets in Eq. \eqref{Hps} is known as the topological term and in the absence of other terms it causes decoherence of the state of the principal system without any dissipation of energy. This {\it decoherence without dissipation} is considered to be one of the distinctive features of the spin bath, as opposed to the oscillator bath \cite{Prokofev93, Prokofev2000}. We, however, show in Sec. \ref{OscillatorBath} that such effect exists in the oscillator bath model as well. 

The tunneling matrix element $\Delta$ is typically the smallest energy scale in Hamiltonian \eqref{Hps}. A typical example of the system and environment that lead to Hamiltonian \eqref{Hps} is a single molecule magnet (such as Fe8, Mn12, $\cdots$) and its surrounding nuclear spins and defects. The single molecule magnet forms a double well energy landscape with a very small tunnel splitting (e.g. for Fe8: $\Delta \sim 0.1 \ \mu K$, $V_{k,k'}^{\alpha \beta} \sim 1 \ \mu K$, $\xi \sim 0.1 \ mK$, and $\xi_0 \sim 10 \ mK$) \cite{Prokofev1998, Prokofev93, Tupitsyn97, Prokofev2000, Prokofev95, Prokofev95b, Gatteschi2006, Garg2013, Garg2015, Wernsdorfer1999, Garg2009}. 

In the strong coupling limit, the solution of Hamiltonian \eqref{Hps} under most conditions, and for practical cases (such as Fe8 and Mn12 as the qubit \cite{Prokofev1998}),  is an incoherent relaxation of the principal system with relaxation rate, up to a factor of order unity \cite{Prokofev1998, Prokofev2000},
\beq \label{Gamma}
\Gamma(\xi) \approx \f{\Delta^2}{\xi_0} e^ {-\abs{\f{\xi}{\xi_0}}}.
\eeq
As one can see $\Gamma(\xi)$ is a decreasing function of $\xi$. As $\xi$ becomes significantly larger than $\xi_0$ the relaxation becomes significantly small and the system can hardly tunnel from one well to the other well. Thus, as long as the bias energy of the system $\xi$ is  smaller than the width of the fluctuating field of the spin bath in energy unit $\xi_0$, the system will tunnel, otherwise it will remain in one well.  

We note that in some cases of theoretical interest the relaxation rate is smaller than that of Eq. \eqref{Gamma} and is given by 
\beq \label{Gamma'}
\Gamma(\xi) \approx \f{\Delta^2}{\Gamma_2} e^ {-\abs{\f{\xi}{\xi_0}}}
\eeq
where $\Gamma_2 \ge \xi_0$ and depends on the parameters of the spin bath \cite{Prokofev1998, Prokofev2000}. 

In next section we turn to oscillator bath model and give a brief review of that. Then, in Sec. \ref{Sim} we replace the spin bath, for the problem of interaction of a qubit with an environment, with the oscillator bath and choose the parameters of the oscillator bath such that it simulates the effect of the spin bath and produces relaxation rates similar to that of Eqs. (\ref{Gamma}-\ref{Gamma'}). 


\section{Oscillator Bath Model} \label{OscillatorBath}
The oscillator bath model studies the effect of an environment made of simple harmonic oscillators on a principal system. The self Hamiltonian of the environment, which is called the oscillator bath, is  
\beq
H_{B,osc} = \sum_i \f{\hat{p}_i^2}{2 m_i} + \f{1}{2} m_i \omega_i^2 \hat{x}_i^2
\eeq
where $m_i$, $\omega_i$, $\hat{x}_i$, and $\hat{p}_i$, are mass, frequency, position and momentum operators of the ith oscillator in the bath, respectively. Many forms of  interaction of the oscillators with a truncated two-state system can be cast into the following form \cite{Caldeira83, Leggett1984, Leggett87}, known as the spin-boson Hamiltonian, 
\beqn \label{HUosc}
\nn H_{U,osc} &=& -\f{\Delta}{2} \ \hat{\tau}_x - \f{\xi}{2} \ \hat{\tau}_z + \f{\hat{\tau}_z}{2} \sum_i c_i \ \hat{x}_i \\ 
&+&\sum_i \f{\hat{p}_i^2}{2 m_i} + \f{1}{2} m_i \omega_i^2 \hat{x}_i^2
\eeqn
The dynamics of the principal system (the qubit) under influence of the above oscillator bath in various regimes of parameters has been investigated by Leggett, his collaborators,  and many other authors in the literature \cite{Leggett87, Caldeira83, Leggett1984, Weiss2008, Nesi2007, Thorwart2004, Nesi2007b, Grifoni1999}. In the regime that the bath has a continuum of low frequency modes, i.e. $\omega_i \lesssim \Delta$ for an infinite number of oscillators in the bath, the dynamics of the qubit is dissipative and irreversible. Dissipation causes transfer of energy from the qubit to the environment. That in turn gives rise to relaxation of the state of the qubit from excited states to the ground state ($T_1$ relaxation) as well as dephasing of the state of the qubit ($T_2$ relaxation). 

For nonzero bias energy $\xi \ne 0$ in Hamiltonian \eqref{HUosc}, there is, however,  another contribution to dephasing known as pure dephasing ($\tau_\phi$ process). Pure dephasing does not transfer any energy between the qubit and the environment. To isolate the term responsible for pure dephasing one can write Hamiltonian \eqref{HUosc} in the basis that diagonalizes the self Hamiltonian of the qubit as follows
\beqn \label{HUosc2}
\nn H_{U,osc} &=& -\f{\Delta E}{2} \ \hat{\rho}_z + \f{1}{2}(\sin \eta \ \hat{\rho}_x + \cos \eta \  \hat{\rho}_z) \sum_i c_i \ \hat{x}_i \\ 
&+&\sum_i \f{\hat{p}_i^2}{2 m_i} + \f{1}{2} m_i \omega_i^2 \hat{x}_i^2
\eeqn
where $\Delta E = (\Delta^2 + \xi^2)^{1/2}$, $\cot \eta = \xi/\Delta$, and $\vec{\hat{\rho}}$ are Pauli matrices in the new basis.  The term $\cos \eta \  \hat{\rho}_z \sum_i c_i \hat{x}_i$ in Eq. \eqref{HUosc2} causes pure dephasing while the term $\sin \eta \  \hat{\rho}_x \sum_i c_i \hat{x}_i$ only causes dissipation and its resultant dephasing, together.  

Therefore, pure dephasing effect is not a distinctive feature of spin bath, but it exists in oscillator bath as well (cf. \cite{Prokofev93, Prokofev2000}). Moreover, the oscillator bath can be coupled to the qubit through any component of the spin of the qubit (See  Sec. IV.C of \cite{Makhlin2001}). The interaction term in Eq. \eqref{HUosc}, for instance, can be replaced with $\f{\hat{\tau}_x}{2} \sum_i c_i \ \hat{x}_i$. In this case when the bias energy vanishes, $\xi=0$, the effect of the oscillator bath is only pure dephasing and decoherence takes place without any dissipation.  

Here, we use the spin-boson Hamiltonian in the form of Eq. \eqref{HUosc}. We are interested in the solution of Eq. \eqref{HUosc} in the small $\Delta$ limit when $\Delta \ll \xi$.  This is the regime that we can relate to the spin bath model in the strong coupling limit since $\Delta$ is the smallest energy scale in the spin bath model in that limit. In the regime of $\Delta \ll \xi$, one can do perturbation theory in $\Delta$ to solve Eq. (\ref{HUosc2}) for the dynamics of the qubit. This method is known as {\it golden rule} and is explained in details in Ref. \cite{Leggett87}. Here we quote the result: The resultant dynamics of the qubit in this regime is an {\it incoherent relaxation} with relaxation rate 
\beq \label{tau} 
\Gamma(\xi) = \Delta^2 \int_{0}^{\infty} dt \ \cos(\xi \ t) \ \cos(\f{Q_1(t)}{\pi}) \ e^{-Q_2(t)/\pi}
\eeq 
where 
\begin{align} \label{Q1}
Q_1(t) &= \int_{0}^{\infty} \f{J(\omega)}{\omega^2} \sin(\omega t) d\omega, \\ \label{Q2}
Q_2(t) &= \int_{0}^{\infty} \f{J(\omega)}{\omega^2} (1 - \cos(\omega t)) \coth(\omega/2 k T) \ d\omega,
\end{align}
and 
\beq \label{JJ}
J(\omega) = \f{\pi}{2} \sum_i \f{c_i^2}{m_i \omega_i} \delta(\omega - \omega_i),
\eeq
is the spectral density function of the oscillator bath \cite{Caldeira83, Leggett87}. In order to simulate the effect of the spin bath by the oscillator bath, we choose, in the next section, $J(\omega)$ and temperature of the oscillator bath such that it produces the relaxation rates of the spin bath, Eq. (\ref{Gamma}-\ref{Gamma'}). 

\section{Simulation of Spin bath by oscillator bath in strong coupling limit} \label{Sim}
We choose an oscillator bath at zero temperature with spectral density function 
\beq 
\label{J}
J(\omega) = 2 \pi \alpha \ \omega \ e^{-\omega/ \xi_0}
\eeq
where $\alpha = 1/2$ and $\xi_0$ is the width of the bias field fluctuation of the original spin bath, in unit of energy, that we want to simulate its effect (see Sec. \ref{SpinBath}). Our goal here is to {\it simulate} the effect of the spin bath in the strong coupling limit by an oscillator bath. The simulator oscillator bath may not bear all the physical features that the spin bath has. However, the effect that it produces is similar to that of the spin bath. In addition, as we see shortly, the important feature of the above oscillator bath, which is the width of the bias energy fluctuation, is comparable to that of the spin bath. Finally, we  note that our choice for the spectral density function and temperature may not be unique. There may well be other choices that produce similar effects and one may be interested in finding them.  

\begin{figure}[htbp]
\begin{center}
\includegraphics[width=0.5\textwidth]{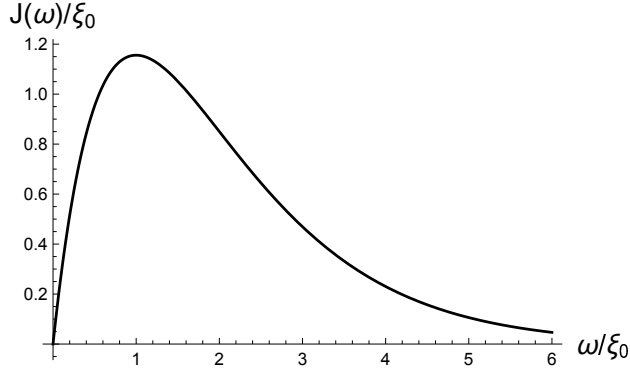}
\caption{Spectral density function $J(\omega)$ of the simulator oscillator bath of the spin bath, Eq. \eqref{J}. $\xi_0$ is the width of fluctuation of the bias energy of the spin bath. }
\label{Fig3}
\end{center}
\end{figure}

The spectral density function of Eq. \eqref{J} is ohmic at small frequencies $\omega \ll \xi_0$, peaks at $\omega = \xi_0$ and then falls off and asymptotes to zero at large frequencies $\omega \gg \xi_0$ as shown in Fig. \ref{Fig3}. 

The spectral density function can be used to estimate the width of the fluctuation of the bias energy. To this end, we first note that the spin-boson Hamiltonian \eqref{HUosc} can be written as 
\beqn \label{HUosc2'}
\nn H_{U,osc} &=& -\f{\Delta}{2} \ \hat{\tau}_x + \f{\hat{\xi}_B - \xi}{2} \ \hat{\tau}_z + H_{B,osc}
\eeqn
where 
\beq
\hat{\xi}_B = \sum_i c_i \hat{x}_i
\eeq
is the bias energy operator of the bath. The width of fluctuation of the bias energy is evidently $\sqrt{\langle \xi_B^2 \rangle}$. For an oscillator bath at zero temperature, one can estimate $\langle \xi_B^2 \rangle$ as follows
\beqn
\nn \langle \xi_B^2 \rangle &=& \langle \sum_{i,j} c_i c_j x_i x_j \rangle \\  \label{xiB2}
&\simeq& \langle \sum_{i} c_i^2 x_i^2 \rangle \simeq \sum_i \f{c_i^2}{m_i \omega_i}
\eeqn 
where we neglected the cross terms $\langle x_i x_j \rangle$, used the ground state values for $\langle x_i^2 \rangle$, and set $\hbar = 1$ as before. The right hand side of  \eqref{xiB2} can be written in terms of the spectral density function \eqref{JJ},
\beq \label{cJ}
\sum_i \f{c_i^2}{m_i \omega_i} = \f{2}{\pi} \int_0^{\infty} J(\omega) d \omega.
\eeq
From \eqref{xiB2} and \eqref{cJ} we deduce 
\beq \label{Jxi}
\langle \xi_B^2 \rangle \simeq \f{2}{\pi} \int_0^{\infty} J(\omega) d \omega.
\eeq
We, hence, substitute our simulator bath spectral density function \eqref{J} into Eq. \eqref{Jxi} to obtain  
\beq \label{Jxii}
\langle \xi_B^2 \rangle = 2 \xi_0^2.
\eeq
Eq. \eqref{Jxii} indicates that the width of the fluctuation of the bias energy in the spin bath model, $\xi_0$, is of the same order of magnitude as that of its simulator oscillator bath 
\beq
\sqrt{\langle \xi_B^2 \rangle} \sim 1.4 \xi_0.
\eeq
Therefore, the two baths have this important feature in common. 

Finally, we calculate the relaxation rate of the simulator oscillator bath. We substitute the spectral density function \eqref{J} and zero temperature of the oscillator bath into Eqs. (\ref{Q1}-\ref{Q2}) to obtain 
\beqn \label{Q1t}
Q_1(t) &=& 2 \ \pi \alpha \ \tan^{-1} \xi_0 t \\ \label{Q2t}
Q_2(t) &=& \alpha \ \pi  \ \ln (1 + \xi_0^2 t^2).
\eeqn
The above results for $Q_1(t)$ and $Q_2(t)$ are the same that are obtained in Ref. \cite{Leggett87} for an ohmic oscillator bath with a cut off frequency when one sets the cutoff frequency to $\xi_0$ and temperature to zero (See Eq. (5.4) in Ref. \cite{Leggett87}). We, next, substitute $Q_1(t)$ and $Q_2(t)$ from Eqs. (\ref{Q1t}-\ref{Q2t}), for the value of $\alpha = 1/2$, into Eq. \eqref{tau} and take the integral. The integration is performed in the Supplementary material. The result is as follows
\beq \label{Gamma2}
\Gamma(\xi) = \f{\pi \Delta^2}{2 \xi_0}  e^{-\abs{\f{\xi}{\xi_0}}}. 
\eeq
This relaxation rate has the same form as that of the spin bath in Eq. \eqref{Gamma}, up to a factor of order unity. 

To simulate the relaxation rate of Eq. \eqref{Gamma'} one needs to add a set of high frequency oscillators to the oscillator bath in Eq. \eqref{HUosc}. The high frequency oscillators do not affect the dynamics of the qubit, however, they renormalize the tunneling matrix element to a smaller effective value $\Delta_{\text{eff}}$. This effect of high frequency oscillators has been discussed in details in Sec. II of Ref. \cite{Leggett87}. The qubit under the influence of the new bath, which is the previous bath plus a set of high frequency oscillators appended to it, relaxes incoherently, as before. However, the relaxation rate now becomes  
\beq \label{Gamma3}
\Gamma(\xi) = \f{\pi \Delta_{\text{eff}}^2}{2 \xi_0}  e^{-\abs{\f{\xi}{\xi_0}}}. 
\eeq
One can choose the spectral density function of the high frequency oscillators $J'(\omega)$ such that 
\beq
\f{\Delta_{\text{eff}}^2}{\xi_0} = \f{\Delta^2}{\Gamma_2}.
\eeq
Thus, one obtains for the relaxation rate
\beq \label{Gamma4}
\Gamma(\xi) = \f{\pi \Delta^2}{2 \Gamma_2}  e^{-\abs{\f{\xi}{\xi_0}}} 
\eeq
that is the same as the relaxation rate of the spin bath in Eq. \eqref{Gamma'}, up to a factor of order unity. One may fix the numerical factors by further tuning $J'(\omega)$ as well. 

Relaxation rates of \eqref{Gamma2} and \eqref{Gamma4} are decreasing functions of the bias energy $\xi$ just like in the spin bath case. They also have the same form and match with those of spin bath. Therefore, we conclude that the oscillator baths that we chose in this section can simulate the effect of the spin baths in the strong coupling limit of the spin baths. 

\section{Conclusion} \label{con}
We have shown in this paper that an oscillator bath can simulate the effect of the spin bath in the strong coupling limit of the spin bath. This is the limit that has been thought to have strikingly different effects on principal systems. We showed, however, that by choosing appropriate values for the spectral density function of the oscillator bath and its temperature, the oscillator bath can simulate the effect of the spin bath. 

The nature of the two baths are different. In the spin bath, the spins may interact with one another, while in the oscillator bath the oscillators are non-interacting. Each spin in the spin bath, individually, has a few number of excited states, while each oscillator in the oscillator bath has an infinite number of excited states available to it. However, the interest in modeling quantum environments is for the effect that they may exert on a system, not the internal structures of them. We demonstrated that in spite of the differences between the spin bath and oscillator bath, the overall effect of the two baths on qubits can be similar in the strong coupling limit of the spin bath. 

We have also shown that the oscillator bath can cause decoherence without dissipation just like the spin bath. 

The result of this paper indicates that for most practical purposes the difference between the spin bath and the oscillator bath, as far as the relaxation rate is  concerned, is far less than has been previously recognized. It will be interesting to find how the result of this paper fits in the broader picture of simulation of quantum noise by classical noise \cite{Crow2014,Halataei2017QNsim}.


\section*{Acknowledgement}
I am indebted to my PhD adviser, Anthony J. Leggett, for the stimulating and fruitful discussions and his support. I am grateful to Mahdieh Piran for her permanent support as well. 


\appendix
\setcounter{secnumdepth}{0} 
\section{Supplementary material} \label{app}
Here, we perform the integration in Eq. (\ref{tau}) for $Q_1(t)$ and $Q_2(t)$ of Eqs. (\ref{Q1t}-\ref{Q2t}) when $\alpha = 1/2$. 

First, we substitute $Q_1(t)$ and $Q_2(t)$ from Eqs. (\ref{Q1t}-\ref{Q2t}) into cosine and exponential functions, respectively, to get,
\beqn
\cos (Q_1(t)/\pi) &=&  \f{1}{\sqrt{1+ \xi_0^2 t^2}}, \\
\exp (-Q_2(t)/\pi) &=&  \f{1}{\sqrt{1+ \xi_0^2 t^2}}.
\eeqn
Next, we substitute the above results into Eq. (\ref{tau}) 
\beq \label{Gammaxi}
\Gamma(\xi) = \Delta^2 \int_{0}^{\infty} \f{\cos(\xi \ t)}{1 + \xi_0^2 t^2} \ dt . 
\eeq
The integrand in Eq. \eqref{Gammaxi} is an even function of $t$. Therefore, we can extend the integral to $-\infty$ for a prefactor of $1/2$, 
\beq \label{Gammaxi2}
\Gamma(\xi) = \f{\Delta^2}{2} \int_{-\infty}^{\infty} \f{\cos(\xi \ t)}{1 + \xi_0^2 t^2} \ dt . 
\eeq
Now, we use the identity  
\beq
\cos \xi t = \f{e^{i \abs{\xi} t} + e^{-i \abs{\xi} t}}{2}   
\eeq
in Eq. \eqref{Gammaxi2} and split the integral into two parts to prepare them for contour integrations
\beq \label{Gammaxi3}
\Gamma(\xi) = \f{\Delta^2}{4} \int_{-\infty}^{\infty} \f{e^{i \abs{\xi} t}}{1 + \xi_0^2 t^2} \ dt + \f{\Delta^2}{4} \int_{-\infty}^{\infty} \f{e^{-i \abs{\xi} t}}{1 + \xi_0^2 t^2} \ dt 
\eeq
The above integrands have poles at $t = \pm i/ \xi_0$. We close the contour of the first integral in the upper half plane and the second one in the lower half plane. Then we use the residue theorem to obtain
\beqn \label{Gammaxi4}
\nn \Gamma(\xi) &=& \f{\Delta^2}{4} \left\{ 2 \pi i \f{e^{i \abs{\xi} i / \xi_0}}{2 i \xi_0} - 2 \pi i \f{e^{-i \abs{\xi} (-i) / \xi_0}}{-2 i \xi_0} \right\} \\
&=& \f{\pi \Delta^2}{2 \xi_0}  e^{-\abs{\f{\xi}{\xi_0}}}
\eeqn
Eq. \eqref{Gammaxi4} is the result we wanted to obtain and is the same as Eq. \eqref{Gamma2} in the text.

\bibliographystyle{unsrt}
\bibliography{thesisrefs} 

\begin{thebibliography}{10}

\bibitem{Leggett1989}
Anthony~J. Leggett.
\newblock Quantum mechanics of complex systems.
\newblock In Alan R.~Bishop Dionys~Baeriswyl and Jose Carmelo, editors, {\em
  Applications of statistical and field theory methods to condensed matter},
  pages 1--25. Plenum Press, New York and London, 1989.

\bibitem{BohrCollected}
{\em Niels Bohr Collected Works: Vols 2-9}.
\newblock North-Holland Publishing Company, 1981-1986.

\bibitem{Caldeira83}
A.O Caldeira and A.J Leggett.
\newblock Quantum tunnelling in a dissipative system.
\newblock {\em Annals of Physics}, 149(2):374 -- 456, 1983.

\bibitem{Leggett87}
A.~J. Leggett, S.~Chakravarty, A.~T. Dorsey, Matthew P.~A. Fisher, Anupam Garg,
  and W.~Zwerger.
\newblock Dynamics of the dissipative two-state system.
\newblock {\em Rev. Mod. Phys.}, 59:1--85, Jan 1987.

\bibitem{Friedman2000}
Jonathan~R Friedman, Vijay Patel, Wei Chen, SK~Tolpygo, and James~E Lukens.
\newblock Quantum superposition of distinct macroscopic states.
\newblock {\em nature}, 406(6791):43--46, 2000.

\bibitem{vanderWal2000}
Caspar~H Van Der~Wal, ACJ Ter~Haar, FK~Wilhelm, RN~Schouten, CJPM Harmans,
  TP~Orlando, Seth Lloyd, and JE~Mooij.
\newblock Quantum superposition of macroscopic persistent-current states.
\newblock {\em Science}, 290(5492):773--777, 2000.

\bibitem{Makhlin2001}
Yuriy Makhlin, Gerd Sch\"on, and Alexander Shnirman.
\newblock Quantum-state engineering with josephson-junction devices.
\newblock {\em Rev. Mod. Phys.}, 73:357--400, May 2001.

\bibitem{Clarke2008}
John Clarke and Frank~K Wilhelm.
\newblock Superconducting quantum bits.
\newblock {\em Nature}, 453(7198):1031--1042, 2008.

\bibitem{Oliver2011}
Jonas Bylander, Simon Gustavsson, Fei Yan, Fumiki Yoshihara, Khalil Harrabi,
  George Fitch, David~G Cory, Yasunobu Nakamura, Jaw-Shen Tsai, and William~D
  Oliver.
\newblock Noise spectroscopy through dynamical decoupling with a
  superconducting flux qubit.
\newblock {\em Nature Physics}, 7(7):565--570, 2011.

\bibitem{Oliver2016}
Fei Yan, Simon Gustavsson, Archana Kamal, Jeffrey Birenbaum, Adam~P Sears,
  David Hover, Ted~J Gudmundsen, Danna Rosenberg, Gabriel Samach, Steven Weber,
  et~al.
\newblock The flux qubit revisited to enhance coherence and reproducibility.
\newblock {\em Nature communications}, 7:12964, 2016.

\bibitem{Prokofev1998}
N.~V. Prokof'ev and P.~C.~E. Stamp.
\newblock Low-temperature quantum relaxation in a system of magnetic
  nanomolecules.
\newblock {\em Phys. Rev. Lett.}, 80:5794--5797, Jun 1998.

\bibitem{Prokofev93}
N~V Prokof'ev and P~C~E Stamp.
\newblock Giant spins and topological decoherence: a hamiltonian approach.
\newblock {\em Journal of Physics: Condensed Matter}, 5(50):L663, 1993.

\bibitem{Tupitsyn97}
I.~S. Tupitsyn, N.~V. Prokof'ev, and P.~C.~E. Stamp.
\newblock Effective hamiltonian in the problem of a central spin coupled to a
  spin environment.
\newblock {\em International Journal of Modern Physics B}, 11(24):2901--2926,
  1997.

\bibitem{Prokofev2000}
N~V Prokof'ev and P~C~E Stamp.
\newblock Theory of the spin bath.
\newblock {\em Reports on Progress in Physics}, 63(4):669, 2000.

\bibitem{Prokofev95}
N.V. Prokofev and P.C.E. Stamp.
\newblock Spin environments and the suppression of quantum coherence.
\newblock In Leon Gunther and Bernard Barbara, editors, {\em Quantum Tunneling
  of Magnetization QTM 94}, volume 301 of {\em NATO ASI Series}, pages
  347--371. Springer Netherlands, 1995.

\bibitem{Prokofev95b}
N.~V. {Prokof'ev} and P.~C.~E. {Stamp}.
\newblock {Decoherence in the Quantum Dynamics of a Central Spin Coupled to a
  Spin Environment}.
\newblock {\em eprint arXiv:cond-mat/9511011}, November 1995.

\bibitem{Schlosshauer2007}
Maximilian~A Schlosshauer.
\newblock {\em Decoherence: and the quantum-to-classical transition}.
\newblock Springer, 2007.

\bibitem{Weiss2008}
Ulrich Weiss.
\newblock {\em Quantum Disipative Systems}.
\newblock World Scientific, 3 edition, 2008.

\bibitem{Nesi2007}
Francesco Nesi, Elisabetta Paladino, Michael Thorwart, and Milena Grifoni.
\newblock Spin-boson dynamics beyond conventional perturbation theories.
\newblock {\em Phys. Rev. B}, 76:155323, Oct 2007.

\bibitem{Thorwart2004}
M~Thorwart, E~Paladino, and M~Grifoni.
\newblock Dynamics of the spin-boson model with a structured environment.
\newblock {\em Chemical Physics}, 296(2):333 -- 344, 2004.
\newblock The Spin-Boson Problem: From Electron Transfer to Quantum Computing
  ... to the 60th Birthday of Professor Ulrich Weiss.

\bibitem{Grifoni1999}
M.~Grifoni, E.~Paladino, and U.~Weiss.
\newblock Dissipation, decoherence and preparation effects in the spin-boson
  system.
\newblock {\em The European Physical Journal B - Condensed Matter and Complex
  Systems}, 10(4):719--729, Jun 1999.

\bibitem{Nesi2007b}
F.~Nesi, E.~Paladino, M.~Thorwart, and M.~Grifoni.
\newblock Spin-boson dynamics: A unified approach from weak to strong coupling.
\newblock {\em EPL (Europhysics Letters)}, 80(4):40005, 2007.

\bibitem{Eckern84}
Ulrich Eckern, Gerd Sch\"on, and Vinay Ambegaokar.
\newblock Quantum dynamics of a superconducting tunnel junction.
\newblock {\em Phys. Rev. B}, 30:6419--6431, Dec 1984.

\bibitem{Gatteschi2006}
Dante Gatteschi, Roberta Sessoli, and Jacques Villain.
\newblock {\em Molecular nanomagnets}, volume~5.
\newblock Oxford University Press on Demand, 2006.

\bibitem{Garg2013}
Avinash Vijayaraghavan and Anupam Garg.
\newblock Low-temperature magnetization relaxation in magnetic molecular
  solids.
\newblock {\em Annals of Physics}, 335:1 -- 20, 2013.

\bibitem{Garg2015}
Erik Lenferink, Avinash Vijayaraghavan, and Anupam Garg.
\newblock Low-temperature magnetization dynamics of magnetic molecular solids
  in a swept field.
\newblock {\em Annals of Physics}, 356:37 -- 56, 2015.

\bibitem{Wernsdorfer1999}
W.~Wernsdorfer and R.~Sessoli.
\newblock Quantum phase interference and parity effects in magnetic molecular
  clusters.
\newblock {\em Science}, 284(5411):133--135, 1999.

\bibitem{Takahashi2011}
S~Takahashi, IS~Tupitsyn, J~Van~Tol, CC~Beedle, DN~Hendrickson, and PCE Stamp.
\newblock Decoherence in crystals of quantum molecular magnets.
\newblock {\em Nature}, 476(7358):76--79, 2011.

\bibitem{Wernsdorfer2002}
W.~Wernsdorfer, S.~Bhaduri, C.~Boskovic, G.~Christou, and D.~N. Hendrickson.
\newblock Spin-parity dependent tunneling of magnetization in single-molecule
  magnets.
\newblock {\em Phys. Rev. B}, 65:180403, Apr 2002.

\bibitem{Garg2009}
Avinash Vijayaraghavan and Anupam Garg.
\newblock Incoherent landau-zener-st\"uckelberg transitions in single-molecule
  magnets.
\newblock {\em Phys. Rev. B}, 79:104423, Mar 2009.

\bibitem{Caldeira93}
A.~O. Caldeira, A.~H. Castro~Neto, and T.~Oliveira~de Carvalho.
\newblock Dissipative quantum systems modeled by a two-level-reservoir
  coupling.
\newblock {\em Phys. Rev. B}, 48:13974--13976, Nov 1993.

\bibitem{energycomment}
We use the term {\it energy landscape} to refer to the classical energy of the
  system when the Hamiltonian is a spin Hamiltonian. In such cases the quantity
  of interest is the classical energy of the system. For example for a single
  large spin with Hamiltonian $H = K_z \hat{S}_z^2 + K_y \hat{S}_y^2$, where
  $\hat{\vec{S}}$ are spin operators and $K_z > K_y > 0$ are constant
  coefficients, the classical Hamiltonian or the energy landscape is $E(\theta,
  \phi) = K_z \cos^2 \theta + K_y \sin^2 \theta \sin^2 \phi$ (See e.g.
  \cite{Chudnovsky88, Loss92, Tupitsyn97}).

\bibitem{xicomment}
$\xi$ is the difference between the minima of the energy landscape in the case
  of spin Hamiltonian \cite{Tupitsyn97, Prokofev2000}. In the case of particle
  Hamiltonian where we use the potential energy, $\xi$ is the difference
  between the energies of the localized ground states of the two wells, in
  absence of tunneling (See e.g. \cite{Leggett87, Halataei2017doublewell}). For
  our problem of interest the formulation of the problem and the result are the
  same in both cases. We, therefore, use the convention of the spin
  Hamiltonians in this paper.

\bibitem{Leggett1984}
A.~J. Leggett.
\newblock Quantum tunneling in the presence of an arbitrary linear dissipation
  mechanism.
\newblock {\em Phys. Rev. B}, 30:1208--1218, Aug 1984.

\bibitem{Crow2014}
Daniel Crow and Robert Joynt.
\newblock Classical simulation of quantum dephasing and depolarizing noise.
\newblock {\em Phys. Rev. A}, 89:042123, Apr 2014.

\bibitem{Halataei2017QNsim}
Seyyed M.~H. Halataei.
\newblock Classical simulation of arbitrary quantum noise.
\newblock {\em Phys. Rev. A}, 96:042338, Oct 2017.

\bibitem{Chudnovsky88}
E.~M. Chudnovsky and L.~Gunther.
\newblock Quantum tunneling of magnetization in small ferromagnetic particles.
\newblock {\em Phys. Rev. Lett.}, 60:661--664, Feb 1988.

\bibitem{Loss92}
Daniel Loss, David~P. DiVincenzo, and G.~Grinstein.
\newblock Suppression of tunneling by interference in half-integer-spin
  particles.
\newblock {\em Phys. Rev. Lett.}, 69:3232--3235, Nov 1992.

\bibitem{Halataei2017doublewell}
Seyyed~MH Halataei and Anthony~J Leggett.
\newblock Tunnel splitting in asymmetric double well potentials: An improved
  wkb calculation.
\newblock {\em arXiv:1703.05758}, 2017.

\end{thebibliography}

\end{document}